\def   \ni {\noindent}

\def   \ssk {\vskip  5truept}

\def   \bsk {\vskip 15truept}

\def   \newline {\hfil\break}

\def\teq#1{$\, #1\,$}
{\catcode`\@=11                                                  
\gdef\SchlangeUnter#1#2{\lower2pt\vbox{\baselineskip 0pt\lineskip0pt    
\ialign{$\m@th#1\hfil##\hfil$\crcr#2\crcr\sim\crcr}}}}           
\def\gtrsim{\mathrel{\mathpalette\SchlangeUnter>}}               
\def\lesssim{\mathrel{\mathpalette\SchlangeUnter<}}   
\documentstyle[epsfig]{article}
\begin{document}

\newcommand{\vol}[2]{$\,$\rm #1\rm , #2.}                 

\pagestyle{empty}

\hsize 5truein
\vsize 8truein
\font\abstract=cmr8
\font\keywords=cmr8
\font\caption=cmr8
\font\references=cmr8
\font\text=cmr10
\font\affiliation=cmssi10
\font\author=cmss10
\font\mc=cmss8
\font\title=cmssbx10 scaled\magstep2
\font\alcit=cmti7 scaled\magstephalf
\font\alcin=cmr6 
\font\ita=cmti8
\font\mma=cmr8
\def\ref{\par\noindent\hangindent 15pt}
\null


\title{\ni 
HIGHLY-MAGNETIZED PULSARS AND INTEGRAL
}                                               

\bsk \bsk
\author{\ni Matthew G. Baring$^{1,2}$ \& Alice K. Harding$^1$}                                                       
\bsk
\affiliation{ 
1) LHEA, NASA Goddard Space Flight Center,
Greenbelt, MD 20771, U.S.A.\\
\ni\vphantom{p}\hskip 0.5truecm
2) Universities Space Research Association, 
Email: baring@lheavx.gsfc.nasa.gov
}                                                
\bsk
\baselineskip = 12pt

\abstract{ABSTRACT \ni
The complete absence of radio pulsars with periods exceeding a few
seconds has lead to the popular notion of the existence of a high
period death line.  We have recently postulated the existence of
another radio quiescence boundary at high magnetic fields
(\teq{B\gtrsim 4\times 10^{13}}G) in the upper portion of the
period-period derivative diagram, a domain where no radio pulsars are
observed.  The origin of this high B boundary is also due to the
suppression of magnetic pair creation, \teq{\gamma\to e^{\pm}}, but
mainly because of competition with the exotic QED process of magnetic
photon splitting, \teq{\gamma\to\gamma\gamma}, coupled with ground
state pair creation.  This mechanism could also explain the low
spectral cutoff energy of the gamma-ray pulsar PSR1509-58, which lies
near the high B death-line.  In this paper, we summarize the hypothesis
of this new ``death line,'' and discuss some subtleties of pair
suppression that relate to photon polarization and positronium
formation.  We identify several ways in which Integral will serve as a
useful tool in probing high field pulsars, given that the radio
quiescence boundary is transparent to the high energy bands.
}                                                    
\bsk
\baselineskip = 12pt

\bsk
\baselineskip = 12pt


\text{
\ni 1. HIGHLY-MAGNETIZED PULSARS\ssk
\ni     

The study of pulsars with unusually high magnetic fields, namely
\teq{B\gtrsim 10^{13}}G, has recently become of great interest in the
astronomical community, due both to the rapid increase in
observational data indicating such high fields, and also to the
fascinating physics that might arise in their environs.  Apart from a
handful of conventional radio pulsars such as PSR 1509-58 with
spin-down field estimates in the range \teq{10^{13}}G\teq{\lesssim
B\lesssim 3\times 10^{13}}G, there is the growing body of anomalous
X-ray pulsars (AXPs) and soft gamma repeaters (SGRs) perhaps with much
larger fields.  Such sources are currently being touted as candidate
{\it magnetars} (e.g. Thompson \& Duncan 1993), a class of neutron
stars with fields in excess of \teq{10^{14}}G.

The AXPs are a group of six or seven pulsating X-ray sources with
periods around 6-12 seconds, which are anomalous in comparison with
average characteristics of known accreting X-ray pulsars.  They are
bright, steady X-ray sources having luminosities \teq{L_X \sim
10^{35}\,\rm erg\; s^{-1}}, they show no sign of any companion, are
steadily spinning down, and have ages \teq{\tau \lesssim 10^5} years.
Those which have measured \teq{\dot P} (e.g. Mereghetti \& Stella 1995;
Gotthelf \& Vasisht 1998) have derived dipole spin-down magnetic fields
between \teq{10^{14}} and \teq{10^{15}} Gauss.  The SGRs, so-named
because of repeated transient $\gamma$-ray burst activity, are another
type of high-energy source that has recently joined this group of
possible magnetars.  There are four known SGR sources (one newly
discovered:  Kouveliotou, et al.  1998b), two (SGRs 1806-20 and
0526-66) associated with young (\teq{\tau < 10^5} yr) supernova
remnants.  Recently, 7.47s and 5.16s pulsations have been discovered
(Kouveliotou, et al. 1998a,c; Hurley et al. 1998) in the quiescent
X-ray emission of SGRs 1806-20 and 1900+14, respectively, with SGR
1900+14 exhibiting a 5.15s period in a $\gamma$-ray burst (Cline et
al.  1998), much like the original and canonical 5th March 1979 event
from SGR 0526-66.

In the pulsar \teq{P-\dot P} diagram, both AXPs and SGRs live in a
separate region above the detected radio pulsars:  no radio pulsars
have inferred fields above \teq{\sim 3 \times 10^{13}} Gauss, even
though known selection effects do not {\it a priori} prevent their
detection.  Baring \& Harding (1998a) have recently proposed an
explanation for the absence of radio pulsars of such high
magnetization.  We identified a strong suppression of pair creation,
\teq{\gamma\to e^{\pm}}, in fields above \teq{\sim 10^{13}} Gauss,
ultimately through the action of the exotic QED process of photon
splitting \teq{\gamma\to\gamma\gamma}.  The splitting of one photon
into two lower energy ones (e.g. Adler 1971), which is third-order in
quantum electrodynamics (QED), can act as a competitor to pair creation
\teq{\gamma\to e^+e^-} as an attenuation mechanism for gamma-rays; it
has appreciable reaction rates (e.g. Adler 1971; Baring \& Harding
1997) when \teq{B\gtrsim 10^{13}}Gauss.

The possible relevance of photon splitting for pulsars has recently
received a dramatic boost from the new Comptel data (Kuiper et al.
1998) for PSR 1509-58.  Harding, Baring \& Gonthier (1997, hereafter
HBG97) had discussed \teq{\gamma\to\gamma\gamma} in the context of this
pulsar.  They found that spectral attenuation by splitting at energies
\teq{E} between a few MeV and just below 100 MeV could nicely
accommodate the positive detections by Comptel (then up to around 3
MeV) and the severely constraining EGRET upper limits at \teq{E\gtrsim
100} MeV, provided that polar cap size \teq{\Theta} ranged between
\teq{\Theta\approx 2^\circ} and \teq{\approx 10^\circ}.  Curved
spacetime was incorporated in HBG97's photon propagation calculations,
with splitting (and not pair creation) effecting the predicted spectral
cutoffs.  The positive detection by Kuiper et al. (1998) of pulsed
$\gamma$-rays in the 3--10 MeV band dramatically narrows the available
model phase space.  Spectral cutoffs consistent with the data now
appear only possible for \teq{\Theta\approx 2^\circ}, the standard
polar cap size for PSR 1509-58, and then {\it only provided photon
splitting is acting and general relativity is incorporated in the
model}.  In the absence of splitting, considerably larger \teq{\Theta}
are required to fit the data with attenuation due to \teq{\gamma\to
e^{\pm}}.  Such constraints strongly indicate the action of
\teq{\gamma\to\gamma\gamma} in this source.  Furthermore, the almost
power-law Comptel data suggest that only the
\teq{\perp\to\parallel\parallel} mode of splitting is operating
(discussed in HBG97), a possible experimental verification of
theoretical kinematic selection rules (Adler 1971).

\bsk
\ni 2. RADIO QUIESCENCE \ssk
\ni 

Presuming that a plentiful supply of pairs is a prerequisite for, and
maybe also guarantees, coherent radio emission at observable levels, a
premise of standard polar cap models for radio pulsars, an immediate
consequence of the significant suppression of pair creation by
splitting (and other effects mentioned just below) in pulsars is that
detectable radio fluxes should be strongly inhibited.  Baring \&
Harding (1998a) determined an approximate criterion for the {\it
boundary of radio quiescence} in the \teq{P-\dot P} diagram based on a
comparison of attenuation properties for \teq{\gamma\to e^{\pm}} and
\teq{\gamma\to\gamma\gamma} in general relativistic neutron star
magnetospheres.  We found that for \teq{\dot{P}} above
\teq{\dot{P}\approx 7.9\times 10^{-13}\, (P/1\,\hbox{sec})^{-11/15}},
photon splitting by the \teq{\perp\to\parallel\parallel} mode should
dominate pair creation by \teq{\perp}-polarized photons, corresponding
to fields in the range \teq{3\times 10^{13}}G\teq{\lesssim B\lesssim
8\times 10^{13}}G.  While \teq{\parallel}-polarized photons can still
produce pairs, they are in relative paucity due to the predominance of
\teq{\perp} photons (\teq{\gtrsim 75\%}) in the continuum emission
processes of curvature and synchrotron (or cyclotron) radiation and
resonant Compton upscattering (Baring and Harding 1998b).  Clearly, the
suppression of pair creation by splitting {\it is partial, not total}.

The robustness of this putative boundary for radio quiescence, which is
computed specifically for photon origin near the stellar surface, can
be fully assessed with a Monte Carlo cascade calculation.  Yet, only
moderate overall suppression of pairs per cascade generation is
necessary to quench radio emission given the large pair multiplicities
computed for standard pulsar polar cap cascades (e.g Daugherty \&
Harding 1982).  Baring and Harding (1998b) demonstrate that suppression
of pair creation by splitting is quite effective when the continuum
spectrum is as steep as \teq{E^{-2}}, and/or the maximum photon energy
is lower than around 1 GeV.  They also find that for extended {\it
flat} power-laws (to 10 GeV and above), splitting can actually {\it
enhance} the number of pairs in one generation, principally because it
generates numbers of \teq{\parallel} photons which then have a greater
opportunity to create pairs than their \teq{\perp} counterparts, due to
their lower pair threshold.  Notwithstanding, the \teq{\parallel}
photons produce pairs almost entirely in the ground (0,0) state,
thereby preventing any further pair generations by inhibiting cyclotron
emission.

Our determination of the quiescence boundary is effectively independent
of whether free pair creation or positronium formation is considered
(Baring \& Harding 1998b): the thresholds for bound and free pair
creation are extremely close (Usov \& Melrose, 1995).  The extent of
dissociation of bound pairs has not been decisively determined at
present (see Baring \& Harding 1998b), so it is unclear whether
positronium formation actually inhibits radio signals.  Splitting is
significantly aided in its suppressive action by the dominance of pair
creation in the lowest accessible Landau state configuration when
\teq{B\gtrsim 6\times 10^{12}}Gauss (Harding \& Daugherty 1983).  In
such fields, \teq{\perp}-photons produce pairs no higher than the first
Landau level so that subsequent cyclotron photons are mostly below pair
threshold.  Since \teq{\parallel}-photons leave pairs in the ground
state, they spawn no cyclotron/synchrotron emission.  Hence pair
cascading is strongly inhibited for \teq{B\gtrsim 6\times 10^{12}}G; as
B is increased, \teq{\gamma\to\gamma\gamma} ultimately precipitates a
dramatic reduction in the pair density.  The picture can be modified by
the possibility of resonant Compton scattering by pairs, effecting
polarization mode switching of photons during traversal of the
magnetosphere; this renders pair suppression by splitting more
effective.

\bsk
\ni 3. X-RAY/GAMMA-RAY PROPERTIES AND INTEGRAL \ssk
\ni 

The radio quiescence boundary is transparent to Integral: pulsar X-ray
and gamma-ray fluxes are not expected to be strongly correlated with
the number of pairs.  This is because cascading mechanisms merely
redistribute the emission between the gamma-ray and X-ray bands without
severe diminution of the overall luminosity.  Hence, high field pulsars
should be accessible to Integral.  One of the potential gains from
Integral will be its ability to probe the cyclotronic and
sub-cyclotronic structure in the spectra of high field pulsars.
Integral's continuum and line sensitivities will permit exploration of
spectral bumps and breaks at the cyclotron fundamental.  Furthermore,
given a handful of sources, Integral can search for trends with
spin-down B.  For example, as pair suppression ensues above \teq{\sim
6\times 10^{12}}G, a decline on the number of pair generations and loss
of the steeper synchrotron component to spectra are expected,
corresponding to flatter spectra (case in point, PSR1509-58).  At the
same time, the cyclotron fundamental moves up in energy.  An observed
coupling between such effects would strongly argue in favor of the
polar cap model for high energy pulsar emission.  At the same time, the
issue of the apparent mismatch between Ginga and OSSE spectra for
PSR1509-58 (e.g. see data in HBG97) should be resolved.  In terms of
the AXPs and SGRs, it is unclear what role Integral will play.  It is
possible that Integral will be able to detect or place significant
upper limits to quiescent emission from SGR 1900+14, which has periods
of relatively flat spectra (Kouveliotou et al.  1998c).  Integral
should certainly be capable of detecting any of the SGRs in outburst.
However, extrapolation of the steep RXTE spectra for the quiescent AXPs
falls below the Integral sensitivity.  Hence Integral's impact on
studies of these sources is contingent upon AXPs possessing another
harder component to their emission.


} 

%

\bsk
\baselineskip = 12pt


{\references \ni REFERENCES
\ssk

\def\apj{ApJ}
\def\apjl{ApJ}
\def\nat{Nature}
\def\aap{A\&A}

\rm
\ref 
  Adler, S.~L. 1971, Ann. Phys. \vol{67}{599}
\ref
  Baring, M.~G. \& Harding A.~K. 1997, \apj\vol{482}{372}
\ref
  Baring, M.~G. \& Harding A.~K. 1998a, \apjl, Nov. 1, 1998 issue
  (astro-ph/9809115).
\ref
  Baring, M.~G. \& Harding A.~K. 1998b, in preparation.
\ref
  Cline, T.~L. et al. 1998, IAU Circ. No. 7002.
\ref
  Daugherty, J.~K. \& Harding A.~K. 1982, \apj\vol{252}{337}
\ref
  Gotthelf, E.~V. \& Vasisht, G. 1998, New Astronomy \vol{3}{293}
\ref
  Harding A.~K., Baring, M.~G. \& Gonthier, P.~L. 1997, \apj\vol{476}{246} 
  (HBG97)
\ref
  Harding, A.~K. \& Daugherty, J.~K. 1983, in Positron Electron Pairs in 
  Astrophysics, eds. M.~L. Burns, et al.
  (AIP Conf. Proc. 101: New York), p.~194.
\ref
  Hurley, K. et al. 1998, IAU Circ. No. 7001.
\ref
  Kouveliotou, C. et al. 1998a, \nat\vol{393}{235} 
\ref
  Kouveliotou, C. et al. 1998b, IAU Circ. No. 6944.
\ref
  Kouveliotou, C. et al. 1998c, IAU Circ. No. 7001.
\ref
  Kuiper, L., et al. 1998, these proceedings.
\ref
  Mereghetti, S. \& Stella, L. 1995, \apj\vol{442}{L17}
\ref
  Oosterbroek, T., Parmar, A.~N., Mereghetti, S. \& Israel, G.~L. 1998
     \aap\vol{334}{925}
\ref
  Thompson, C. \& Duncan, R. C. 1993, \apj\vol{408}{194}
\ref
  Usov, V.~V. \& Melrose, D.~B. 1995, Aust. J. Phys. \vol{48}{571}
}                      

\end{document}